\documentclass{andp2012}
\usepackage[english]{babel}
\usepackage{soul}
\setcopyrightyear{2016}
\category{Review Article}

\title{Density correlations and transport in models of many-body localization}

\author{P. Prelov\v{s}ek\inst{1,2}}
\address[1]{Jo\v{z}ef Stefan Institute, SI-1000 Ljubljana, Slovenia}
\address[2]{Faculty of Mathematics and Physics, University of Ljubljana, SI-1000 
Ljubljana, Slovenia}

\author{M. Mierzejewski \inst{3}}
\address[3]{Institute of Physics, University of Silesia, 40-007 Katowice, Poland}

\author{O. Bari\v{s}i\'{c} \inst{4}}
\address[4]{Institute of Physics, HR-1000 Zagreb, Croatia}

\author{J. Herbrych\inst{5,6}}
\address[5]{Department of Physics and Astronomy, The University of Tennessee, Knoxville, 
Tennessee 37996, USA}
\address[6]{Materials Science and Technology Division, Oak Ridge National Laboratory, 
Oak Ridge, Tennessee 37831, USA}

\shortauthors{P. Prelov\v{s}ek et al.}

\begin{abstract}
We present a review of recent theoretical results concerning the many-body localization (MBL) phenomenon, with the emphasis on dynamical density correlations and transport quantities. They are shown to be closely related, providing a comprehensive description of the ergodic-to-nonergodic transition, consistent with experimental findings. While the focus is set mostly on the one-dimensional model of interacting spinless fermions, we also present evidence for the absence of full MBL in the one-dimensional Hubbard model and for the density-wave decay induced by the inter-chain coupling.
\end{abstract}
\shortabstract
\begin{document}
\maketitle
\section{Introduction}
\label{sec:intro}

The idea of many-body localization (MBL) emerged as the extension of the concept of Anderson single-particle localization \cite{anderson58,mott68,kramer93} to interacting many-body (MB) systems \cite{fleishman80,basko06}, in particular to systems of interacting fermions. The MBL physics, its manifestations and open problems related to this novel concept, attracted high attention of theoreticians as well as experimentalists. We give here a short list of some basic hallmarks of the MBL phase in contrast to the 'normal' ergodic phase, ordered more from the perspective of historical appearance and being mostly the result of numerical investigations of concrete models of MBL: a) the Poisson MB level statistics, in contrast to the Wigner-Dyson one in generic 'normal' MB systems \cite{oganesyan07,torres15,luitz15,serbyn16,vasseur16}, b) the logarithmic growth of the entanglement entropy \cite{znidaric08,bardarson12,serbyn15}, as opposed to a linear increase in 'normal' ergodic systems and the saturation in noninteracting (NI) Anderson-localized systems, c) the vanishing of d.c. transport at finite temperatures $T>0$ \cite{berkelbach10,barisic10,agarwal15,lev15,steinigeweg15,barisic16}, d) the nonergodic time evolution of (all) correlation functions, with a related absence of thermalization of initially quenched states \cite{monthus10,pal10,luitz16,mierzejewski16,prelovsek16}, and e) the existence of a full set of local integrals of motion \cite{serbyn13,huse14,rahul15}. We concentrate in this topical review primarily on the aspect of time-dependent density correlations, which are a particular example of ergodic/nonergodic behaviour of correlation functions, as well as on the dynamical and d.c. transport in prototype models of MBL systems.  

Our focus on density correlations and d.c. transport is closely related to experimental efforts to find the MBL physics in cold 
atoms on optical lattices \cite{schreiber15,kondov15,bordia16,choi16}. In one-dimensional (1D) fermionic chains, which simulate the fermionic Hubbard model with potential disorder (in experiment the potential is actually quasiperiodic), the main criterion for the onset of the MBL is the absence of full decay of initial out-of-equilibrium density distribution to the thermal state, i.e., the nonergodic time-evolution of the density imbalance $I(t)$ \cite{schreiber15,bordia16,luschen16}. 
In this context,  $I(t \to \infty) = I_\infty$ can be used as an indicator of the nonergodic (MBL) phase, where 
$I_\infty >0$. On ther other hand, vanishing $I_\infty=0$ can be the signature 
of the 'normal' ergodic phase, whereby one should be aware that the ergodicity condition is much 
more stringent, since it concerns the thermalization of all observables and correlation functions.
The imbalance criterion is used in the experiment of coupled identical 1D chains \cite{bordia16}, where results show that the interchain hopping induces  (at least accelerates) the decay of $I(t)$ along chains. 
$I(t)$ has been recently studied in a bosonic system with a full two-dimensional (2D) disorder as well \cite{choi16}, again with the indication of a well defined onset of $I_\infty >0$. On the other hand, in the disordered three-dimensional (3D) optical lattice of cold fermions the d.c. mobility, i.e., the d.c. velocity $v_{\mathrm{dc}}$ of the gas under the constant force (realized via a gradient of the magnetic field), has been measured \cite{kondov15} in order to locate the transition to the MBL phase with a vanishing d.c. transport.

So far most of theoretical considerations and numerical calculations have been restricted to the 'standard' model of MBL: the disordered 1D model of interacting spinless fermions. Accordingly, we as well predominantly discuss results within this model, as introduced together with its (possibly) experimentally relevant relatives in Sec.~\ref{sec:model}. In Sec.~\ref{sec:dw} we discuss results for the dynamical structure factor and density correlations, together with dynamical and d.c. conductivities in Sec.~\ref{sec:cond}. It has been only recently realized that disordered 1D Hubbard model, which is actually realized in cold-atom experiments, might be qualitatively different. This point is discussed in Sec.~\ref{sec:hubbard}, as well as a particular generalization of coupled identical Hubbard chains. 

\section{Disordered 1D model of interacting spinless fermions}
\label{sec:model}

The 'standard' model for the discussion of the MBL physics is the 1D model of interacting spinless fermions with random (Anderson-type) disorder in local potentials, 
\begin{equation}
H = - t_0 \sum_{i} \left( c^\dagger_{i+1} c_i + c^\dagger_{i} c_{i+1} \right)
+ V \sum_i n_{i+1} n_{i} + \sum_i \epsilon_i n_i\,,
\label{tv}
\end{equation}
where one assumes quenched disorder with the uniform distribution $-W < \epsilon_i <W$, and $W$ is the 
measure of the disorder strength. We use $t_0=1$ as the unit of energy. We note that in the pure limit $W=0$ at zero--temperature and half--filling (average particle density $\bar n=1/2$), the interaction (anisotropy) parameter $\Delta=V/(2 t_0)$ divides the metallic $\Delta < 1$ and the insulating $\Delta \geq 1 $ regime, i.e., the regimes of a weak/modest and a strong interaction.

Via the Jordan-Wigner transformation, the model \eqref{tv} may be mapped on the anisotropic $S=1/2$ Heisenberg model with random local (magnetic) fields $\epsilon_i$,
\begin{equation}
H = J \sum_{i} \left[\frac{1}{2}( S^+_{i+1} S^-_{i} + S^+_i S^-_{i+1})
+ \Delta S^z_{i+1} S^z_{i} \right] + \sum_i \epsilon_i S^z_i\,,
\label{Ham01}
\end{equation}
where $S^z,S^+,S^-$ are standard $S=1/2$ operators and $J= 2t_0$. The spin-chain model \eqref{Ham01} can easily be generalized by introducing random antiferromagnetic exchange interactions $J \to J_i>0$. The latter has a realization in materials for which the low-energy physics is described by local $S=1/2$ spins and randomness emerges from the chemical substitution of different elements, e.g., \mbox{BaCu$_{2}($Si$_{1-x}$Ge$_{x})_{x}$O$_7$} \cite{tsukada99,yamada01,shiroka11,casola12}, Cu$($py$)_{2}($Cl$_{1-x}$Br$_{x})_{2}$ \cite{thede12}, and \mbox{$($Sr$_{1-x}$Ca$_{x})_{2}$CuO$_3$} \cite{Mohan2014}. Such $S=1/2$ systems due to spin-rotation symmetry do not allow for random magnetic fields as in Eq.~\eqref{Ham01}. Still, they exhibit several singular dynamical and static properties \cite{herbrych13,kokalj15}, although probably not yet of localized MBL nature. On the other hand, it has been recently recognized that real materials described by $S=1$ Hamiltonian with single--ion anisotropy, i.e., \mbox{NiCl$_{1-x}$Br$_{x}\cdot4$SC$($NH$_{2})_{2}$} \cite{Yu2012,Wulf2013,Povarov2015}, in the presence of the external magnetic can be mapped on a disordered Heisenberg chain with random magnetic field and can come close to the
MBL physics \cite{herbrych16}. 

\section{Density-wave correlations and dynamical structure factor}
\label{sec:dw}

In this Section we investigate the question whether initially perturbed density of particles does or does not approach the thermal equilibrium. This problem has been studied from two perspectives. One can follow directly the time evolution of a particular initial state (usually a simple ordered one) as performed in the experiments \cite{schreiber15,bordai16} or in numerical simulations \cite{luitz16}. Another one is to study the decay within the linear response theory, i.e., the corresponding equilibrium dynamical (auto)correlation of the relevant operator. The latter represents the relaxation of a weakly perturbed system, averaged over various initial state. The relation between these two approaches will be discussed at the end of the Subsection~\ref{subsec:dcf}.

\subsection{Density correlation functions}
\label{subsec:dcf}

To elucidate the dynamics of out-of-equilibrium density fluctuations, we study the 1D density wave (DW) operator with a general wavevector $q$,
\begin{equation}
n_q = \frac{1}{\sqrt{L}}\sum_{i}\mathrm{e}^{\imath qi} n_{i}\,, 
\end{equation}
and the corresponding equilibrium dynamical susceptibility,
\begin{equation}
\chi(q,\omega)= \imath \int_0^\infty\mathrm{d} t\,
\mathrm{e}^{\imath\omega^+ t} \langle [n_q(t),n_{-q}] \rangle\,,
\label{chiqw}
\end{equation}
where $\omega^+=\omega +\imath \delta$, and $\delta > 0$ is infinitesimal damping (becoming relevant
for a proper analysis of nonergodic systems). $\langle \cdot \rangle$ denotes the average over 
equilibrium states for given $T$.  Such averaging might be experimentally irrelevant  when the system fails to thermally equilibrate and the applicability of equilibrium statistical mechanics looses its background. However, even in such situations,
one might still consider the energy density as the relevant conserved quantity 
and  attribute it to corresponding $T$ which would give the same energy density in 
the thermal equilibrium. Then, the thermal averaging in Eq.~(\ref{chiqw}) represents
a convenient averaging over representative initial states of the system (all states in the  $T \to \infty$ limit) . 
It also allows for some formal steps for correlation functions introduced  furtheron.
In this respect, the approach via the correlation functions differs from
the study of the dynamics of quenched state, which probes just one particular initial wavefunction.  

In a system which might be nonergodic it is convenient to define the related relaxation function,
\begin{equation}
\Phi(q,\omega)= \frac{\chi(q,\omega)-\chi^0(q)}{\omega^+}\,,
\label{phia}
\end{equation}
where $\chi^0(q)$ is the static thermodynamic response, 
\begin{equation}
\chi^0(q)=\int_0^{\beta}\mathrm{d}\tau\,\langle n_q n_{-q}(i \tau) \rangle\,,
\label{chia}
\end{equation}
with $\beta=1/T$ and $\tau$ represents an imaginary time. An ergodic system is characterized by the equality $\chi^0(q)=\chi_q (\omega \to 0)$. However, in a nonergodic system one generally obtains $\chi^0(q) > \chi_q (\omega \to 0)$, so that 
\begin{equation}
\mathrm{Im}~\Phi(q,\omega)= \pi D(q) \delta(\omega) + \phi_{\mathrm{reg}}(q,\omega)\,,
\label{imphia}
\end{equation}
where $D(q)= \chi^0(q) - \chi(q,\omega \to 0)$ is the density stiffness which can be used as an indicator for the onset of the 
MBL phase. On the other hand, $\phi_{\mathrm{reg}}(q,\omega)= \mathrm{Im}\chi(q,\omega)/\omega$ represents the remaining part, which is regular for most systems but might still be singular in the MBL phase. The static susceptibility can be easily evaluated in $\beta\to 0$ limit via Eq.~\eqref{chia}, i.e., $\chi^0(q) = \chi^0 = \beta \bar n (1-\bar n)$ for general average particle density $\bar n$. 

The formalism above is independent of $T$. Since the MBL physics could persist even at high $T$ provided that the disorder $W$ is strong enough, numerical investigations are mostly restricted to the limit $\beta = 1/T \to 0$. Here one can use a simplified relation for the dynamical structure factor
$S(q,\omega)$,
\begin{equation}
S(q,\omega) = \mathrm{Im}\frac{\Phi(q,\omega)}{\pi \beta}=
\frac{1}{\pi} \mathrm{Re} \int_0^\infty\mathrm{d} t\,
\mathrm{e}^{\imath\omega^+ t} \langle n_q(t) n_{-q}  \rangle\,.
\label{sqw}
\end{equation}
While only the staggered wavevector $q=\pi$ has been tested so far in cold-atom systems, investigations of the whole range $q=[0,\pi]$ provides a more complete information. 

Throughout this paper we use the terms ergodic and nonergodic in accordance with the long--time properties of the autocorrelation functions of local operators, as e.g.  in Eq.
(\ref{sqw}).  Whenever  $\langle A(t) A \rangle_{t \rightarrow \infty} \ne 0 $ for some local observable $A$  such that  $\langle  A \rangle =\langle A H \rangle=0$  there exists local or quasilocal conserved quantity
(other than Hamiltonian) which restricts the system dynamics.  Since we study only selected, experimentally  most relevant  local operators,  we may identify only nonergodic  regime, i.e.,
we are able to exclude ergodicity.  The cases where all considered autocorrelation functions vanish in the long-time regime  are possibly ergodic.  
However, we do not exclude the presence of a regime which is intermediate between the ergodic and nonergodic ones \cite{torres16}.

When trying to make direct comparison with the measured imbalance $I(t)$ \cite{schreiber15,bordia16,luschen16}, it should be pointed out that in experimental setup the evolution starts from the uniquely chosen ordered DW state, while our approach at $\beta \rightarrow 0$ introduces averaging over canonical (or microcanonical) ensemble of initial states. In order to explicitly follow the experimental situation, it is enough to replace the trace in Eq.~\eqref{sqw} with a single state $| \psi \rangle$ which should be chosen as the eigenstate $n_q | \psi \rangle=A_q | \psi \rangle $. Then, the correlation function in Eq.~\eqref{sqw} becomes
\begin{equation}
\langle n_q(t) n_{-q}  \rangle \rightarrow A_{-q} 
\langle \psi(t) | n_q | \psi(t) \rangle. 
\end{equation}
While we have formally started from the linear response theory which assumes a small perturbation, the correlation function in Eq.~\eqref{sqw} is not restricted by this condition anymore. The difference is that instead of choosing a single state we get via a (micro)canonical average the information for states in a certain energy window. In the ergodic regime this averaging is plausibly justified by the eigenstate thermalization hypothesis \cite{deutsch91,srednicki94,steinigeweg13} and the typicality concept \cite{steinigeweg14,steinigeweg15}. 

\subsection{Numerical results}

Numerical results for $S(q,\omega)$ \cite{prelovsek16} are obtained by the evaluation of Eq.~\eqref{sqw}, using the microcanonical Lanczos method (MCLM) \cite{long03,prelovsek13}, on finite systems of length $L$ with periodic boundary conditions, and for $\beta \ll 1$. In order to reproduce the most challenging low--$\omega$ regime, and in particular the possible nonergodic contribution, $S(q,\omega\sim0)= S_0(q) \delta(\omega)$ where $S_0(q)=D(q)/\beta$, the MCLM requires large number of Lanczos steps $N_L$. Namely, the $\omega$ resolution is given by $\delta \omega = \Delta E/N_L$, where $\Delta E$ is the many--body eigenenergy span for a given finite-size system.

\begin{figure}[!htb]
\includegraphics[width=\columnwidth]{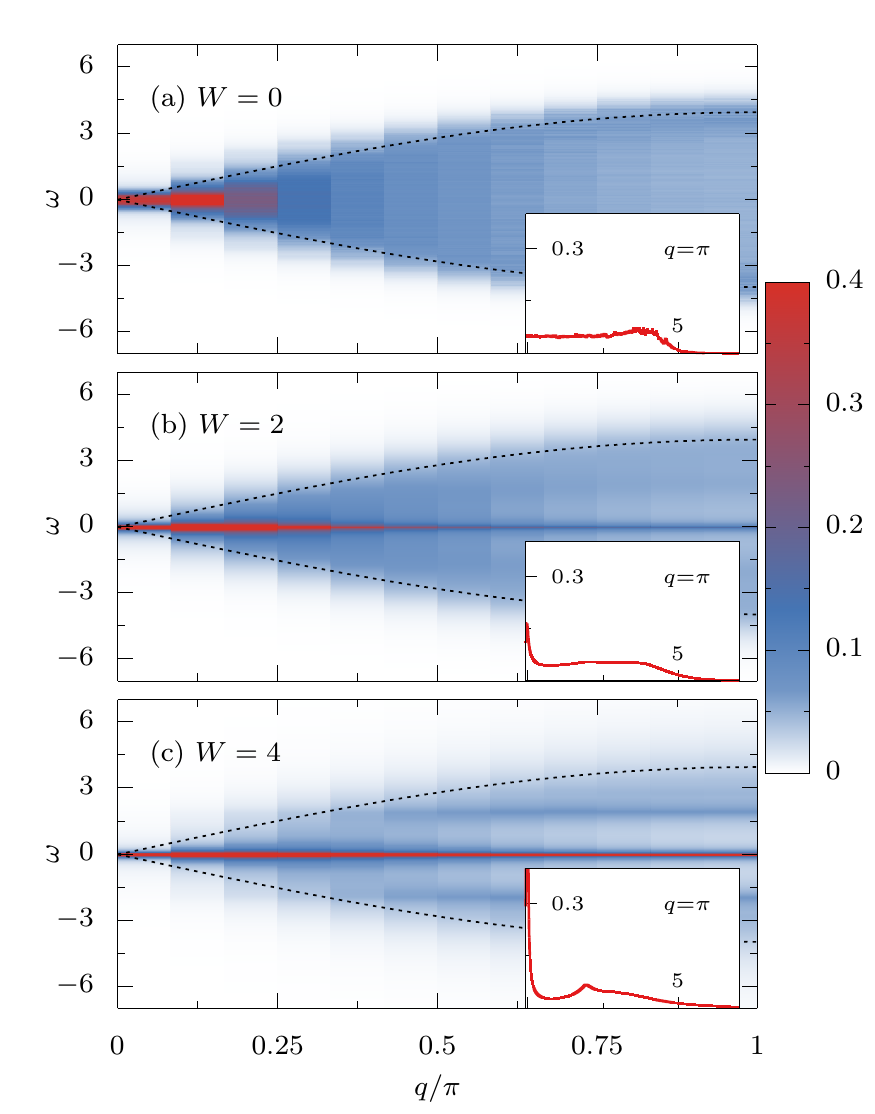}
\caption{\label{fig1}\col Dynamical structure factor $S(q,\omega)$ as calculated within the whole Brillouin zone, 
$q=[0,\pi]$, for $\Delta=1$ on $L=24$ sites, and various disorders $W = 0,2,4$. Results are obtained with MCLM at 
$T \to \infty$. Dashed line: $\Delta(q)=4 \sin(q/2)$. Inset: corresponding $S(q=\pi,\omega)$.}
\end{figure}

In Fig.~\ref{fig1} we present results for $S(q,\omega)$ for the isotropic case $\Delta =1 $ (obtained for $L=24$ sites) averaged over $N_s \sim 100$ random configurations of the disorder $\epsilon_i$ and $N_L=10^4$. We show $S(q,\omega)$ as a color plot in three characteristic regimes of $W$. In the inset of Fig.~\ref{fig1}, the corresponding results for $S(q=\pi,\omega)$ are shown as well:

\noindent a) In the pure case $W=0$, the model is integrable and $S(q,\omega)$ resembles the NI ($V=0$) result \cite{herbrych12} with a renormalized bandwidth $\Delta(q) = 4 t_0 \sin (q/2)$. In particular, at $q>0$ there are 
no additional singularities at $\omega \sim 0$. 
 
\noindent b) At finite but modest $W=2$ we find $S(q,\omega)$ behaving as expected in a generic nonintegrable system. Besides the background similar to the $W=0$ response, there is a well resolved diffusion peak with a width $\delta \omega \propto {\cal D} q^2$ with an apparently small diffusion constant $\cal{D}$ for $W=2$. Still, the maximum at $\omega=0$ persists also for $q=\pi$, which is a precursor of the MBL physics,

\noindent c) For larger $W=4$ having already characteristics of the nonergodic state, the main difference is that instead of the diffusion pole we find for all $q$ a singular contribution $S(q,\omega) = S_0(q) \delta(\omega) +S_{\mathrm{reg}}(q,\omega)$ where $S_0(q) >0 $ is a (density) stiffness indicating the nonergodic phase.

While Fig.~\ref{fig1} gives a rough distinction of different regimes, we provide a finer analysis as well. Following experiments \cite{schreiber15,bordia16,luschen16} we focus here on a particular wavevector $q=\pi$ so that the dynamical structure factor represents the staggered (imbalance) density correlations $C(\omega)= S(q=\pi,\omega)$ at $T \to \infty$. To study the decay, it is particularly informative to present correlations as they develop directly in time, $C(t)$ or in 'quasi-time' $\widetilde{C}(\tau)$,
\begin{equation}
C(t) =  \int_{-\infty }^{\infty} \mathrm{d}\omega \,\mathrm{e}^{-\imath \omega t}
C(\omega)\, ,\quad \widetilde{C}(\tau) = \int_{-1/\tau}^{1/\tau} \mathrm{d}\omega\,C(\omega)\,.
\label{ct}
\end{equation}
\begin{figure}[!htb]
\includegraphics[width=\columnwidth]{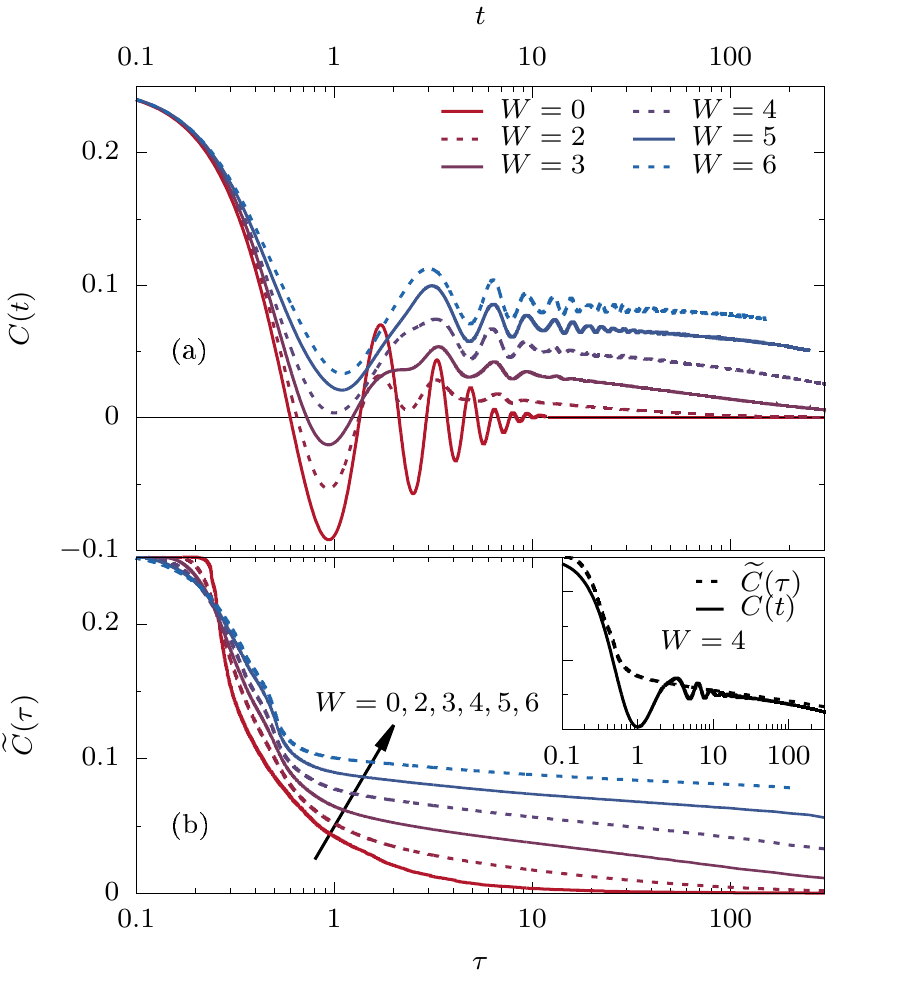}
\caption{\label{fig2} (a) Time evolution of imbalance density correlations $C(t)$ for different disorders $W=0-6$, evaluated at $\Delta=0.5$. (b) The same for $\widetilde C(\tau)$. The inset shows the comparison of both functions at $W=4$. Note the log scale in $t$ and $\tau$. Results are obtained with MCLM on a system of $L=26$ sites and $T \to \infty$.}
\end{figure}
The advantage of $\widetilde{C}(\tau)$ over $C(t)$ is that the former one is a steadily decreasing function in $\tau$, following from positivity $C(\omega) >0$. Both functions are equal at $C(t=0)=\widetilde{C}(\tau \to 0)$, as well as in the limit $t = \tau \to \infty$. Moreover, for  slowly decaying correlation function $C(t) \propto t^{-a}$ one finds the same decay also for $\widetilde{C}(\tau)$, i.e., $ \widetilde{C}(\tau) \propto \tau^{-a}$. In Fig.~\ref{fig2} we present $C(t)$ for $\Delta=0.5$ and for a wide range of disorder $W=[0,6]$, and compare it directly with the 'quasi-time' evolution $\widetilde{C}(\tau)$. 
$C(t)$ displays universal (i.e., independent of disorder $W$) short-time variation , but is at intermediate times $t<10$ dominated by oscillations which have the origin in the noninteracting (NI) model.
 These oscillations represent disorder--averaged dynamics of particles localized on two--site clusters. 
 While the amplitude and the damping of the oscillations depend on  model parameters,  the frequency is determined only by the nearest-neighbor hopping, $\omega^* = 2t_0$, and is correlated with the pronounced peak in optical conductivity 
 $\sigma(\omega)$ (see Fig.~5 and discussion in Sec.~4.1).
 For more details see  \cite{kozarzewski16}. On the other hand, within $\widetilde{C}(\tau)$ such ocillations are 
 filtered out for $\tau > 1$ so the universal decay is better visible, as well as the  final decay either to zero for 
 $W<W^*$ or to $ \widetilde{C}(\tau \rightarrow \infty)=C_0  >0 $ for $W>W^*$.  Fig.~3 shows  $C_0$ vs. $1/L$  as well as disorder--dependence of linearly extrapolated quantity $C^*_0=C_0(1/L \rightarrow 0)$ (where $C_0$ are obtained via exact diagonalization at given $L$). 
\begin{figure}[!htb]
\includegraphics[width=\columnwidth]{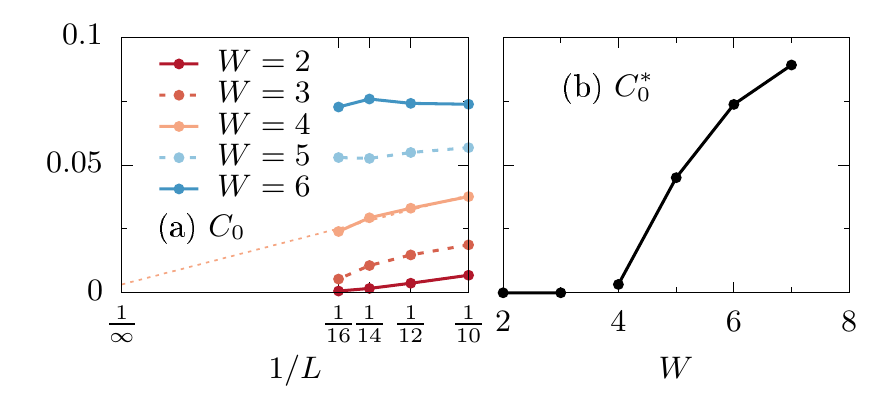}
\caption{\label{fig3} (a) Scaling of density stiffness $C_0$ vs. $1/L$, as obtained for $\Delta=0.5$ via full exact 
diagonalization ($T\to\infty$) for different system sizes $L$. (b) $1/L \to 0$ scaled stiffness $C_0^*$ vs. disorder strength $W$.}
\end{figure}

\begin{figure}[!htb]
\includegraphics[width=\columnwidth]{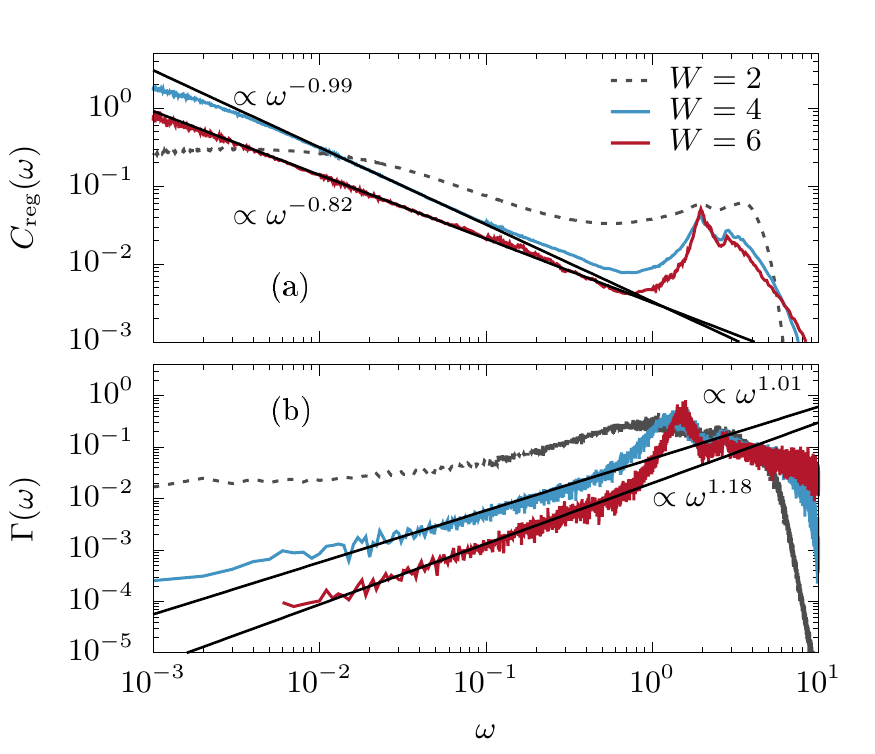}
\caption{\label{fig4} Regular part of (a) density correlations $C_{\mathrm{reg}}(\omega)$ and (b) DW decay rate $\Gamma(\omega)$ vs. $\omega$ for $\Delta=0.5$ and different disorder $W$ (note the log scale), as extracted from MCLM data on $L=26$ sites. Fits to the power law (solid lines) are presented for disorder $W \geq W^* \sim 4$. }
\end{figure}

In the nonergodic regime at $W \geq W^* \sim 4$ the decay is very slow, being of the power-law type 
$\widetilde{C}(\tau)=C_0 + b (\tau/\tau^*)^{-\gamma}$. I
In the limiting case $\gamma \to 0$ this represents a logarithmic variation, 
$\widetilde{C}(\tau) \sim \log(\tau/\tau^*)$. It is therefore informative to analyse MCLM rersults, shown e.g. in Fig. ~2,
using the relation $C(\omega) = C_0 \delta(\omega) +C_{\mathrm{reg}}(\omega)$, with the regular part 
 $C_{\mathrm{reg}}(\omega)$ presented in Fig.~\ref{fig4}a . For $W\geq W^*$ our results are consistent with $C_0 > 0$
and a singular power-law variation $C_{\mathrm{reg}}(\omega) \propto |\omega|^{-\zeta}$, with $0<\zeta\le 1$. 
On the other hand, in the ergodic phase, e.g. for $W=2$, corresponding $C_0=0$ and $C(\omega \to 0)$ saturates,
indicating a 'normal' exponential decay. Still, we  do not exclude more delicate finite-size dependence of the 
latter behaviour which would be possibly consistent with a subdiffusive dynamics \cite{agarwal15,gopal15,
gopal16,luitz16,znidaric16,prelovsek16,luschen16}

\subsection{Memory-function representation}

Since $S(q,\omega)$ is a singular function of $\omega$ at $q \to 0$, and for all $q$ within the MBL phase, it is convenient to employ the memory-function representation. With $\Phi(q,\omega)$ being an analytical function for $\mathrm{Im}\,\omega>0$ \cite{forster95}, one can introduce complex  memory function $M(q,\omega)$, 
\begin{equation}
\Phi(q,\omega)=\frac{- \chi^0(q)}{\omega+M(q,\omega)}, \quad
M(q,\omega) = \imath \frac{g_q^2}{ \chi^0(q) } \widetilde{\sigma} (q,\omega)\,,
\label{mem}
\end{equation} 
related to effective dynamical conductivity $\widetilde{\sigma}(q,\omega)$ via the continuity equation $[H,n_q] = g_q j_q$, where $g_q = 2 \sin(q/2)$ and $j_q$ is the current operator for given $q$. It should be noted that $\widetilde{\sigma}(q,\omega)$ has a straightforward relation to the current correlation function $\sigma(q,\omega)$, as calculated directly for $j_q$, only in the limit $q \to 0$ \cite{forster95}. In this case $\sigma(\omega) = \mathrm{Re}\widetilde{\sigma}(q\to 0, \omega)$ is the optical conductivity with 
the corresponding d.c. value $\sigma_0 = \sigma(\omega=0)$. At finite $q>0$ it follows from the continuity equation that 
$\mathrm{Re}~\sigma(q,\omega \to 0) \propto \omega^2/q^2$, while
$\mathrm{Re}~\widetilde{\sigma}(q, \omega \to 0) >0$ can approach a constant, quite similar as for  $q \to 0$. 

$M(q,\omega)$ and $\widetilde{\sigma}(q,\omega)$ can be calculated using Eqs.~\eqref{mem} from numerically evaluated $S(q,\omega)$. Results obtained for $q=\pi$ \cite{mierzejewski16}, i.e., for the DW damping (imbalance decay rate) $\Gamma(\omega)= \mathrm{Im}~M(q=\pi,\omega) \propto 
\mathrm{Re}~\widetilde{\sigma} (q=\pi,\omega)$ are very similar to (uniform) $q \to 0$ optical conductivity $\sigma(\omega)$, as presented in Fig.~\ref{fig4} and discussed in more detail in Sec.~\ref{sec:cond}: 

\noindent a) The maximum of $\Gamma(\omega)$ for $W\geq 2$ is not at $\omega=0$ (characteristic for simple Drude behaviour), but rather at $\omega^*=2 t_0$.

\noindent b) At $W < W^* \sim 4$ the d.c. limit $\Gamma(\omega \to 0) = \Gamma_0 >0 $ is finite for $W = 2$ 
up to $L =24$, although a vanishing $\Gamma_0$ cannot be excluded for 
larger $L$, particularly in the interval $W* > W > 2$. In the presented case, low-$\omega$ variation 
can be described by $\Gamma(\omega <1) \sim \Gamma_0 + g |\omega|^\alpha$ where $\alpha \lesssim 1$ could be 
an indication for a possible subdiffusion  with $\Gamma_0 \to 0$.

\noindent c) For $W>W^*$ the vanishing of $\Gamma_0$ (although numerical results might be interpreted also with very small but finite $\Gamma_0 < 10^{-3}$) is consistent with the nonergodicity, $C_0 > 0$. Again, we can describe results with a power law $\Gamma(\omega) \propto |\omega|^\alpha$ where $\alpha \gtrsim 1$, and related to $C_{\mathrm{reg}}(\omega) = \omega^{-\zeta}$ with $\alpha = 2- \zeta$ \cite{mierzejewski16}.
As shown in Fig.~\ref{fig4}, we get $\zeta \sim 1$ for $W \sim W^*$ and $\zeta<1$ for 
increasing $W> W^*$.

According to presented results and analysis it is quite plausible that the transition $W = W^*$ to nonergodic (MBL) regime is well characterized with the critical exponent $\alpha=1$ (or equivalently $\zeta=1$), consistent with several other numerical and renormalization-group analysis of dynamical quantities \cite{agarwal15,gopal15,steinigeweg15,potter15,barisic16}. Our results \cite{mierzejewski16} in the nonergodic phase show that the larger the system the longer is the process of the relaxation of $C(t)$. This makes the DW stiffness $C_0>0$ also less evident (at least less reliable) as the well defined
hallmark of the MBL phase, both from numerical calculations as well as from experiments. Hence a more proper definition of the MBL seems to be related just to dynamical critical behaviour.

We can make a further step in the memory-function approach and represent the dynamical conductivity $\widetilde{\sigma}(q,\omega)$ in terms of the current relaxation-rate function $\Lambda(q,\omega)$,
\begin{equation}
\widetilde{\sigma}(q,\omega) = \frac{\imath\chi^0_{j}(q)}{\omega+\Lambda(q,\omega)}\,,
\label{sigqw}
\end{equation}
where $\chi^0_j(q)$ is the static current susceptibility which can be evaluated exactly in the $\beta \to 0$ limit, $\chi^0_{j}(q) = \chi^0_{j} =2 \beta t_0^2 \bar n (1-\bar n)$. We note that such a representation has analogy to the familiar (generalized) Drude formula in the case of $\omega$-independent $\mathrm{Im}~\Lambda(q,\omega \sim 0) \sim \lambda$ so that 
$\lambda=1/\tau_0$ and $\tau_0$ would be the (Drude) current relaxation time. Indeed, this is expected 
to be the case for a weak disorder, where one can employ a perturbative approach to get $\lambda \propto W^2$ \cite{prelovsek16}.
 
Following the perturbation theory valid for weak disorder $W>0$ and interaction $V>0$, it is possible to go beyond the Drude result for $\Lambda(q,\omega)$, in analogy to the theory of dynamical conductivity in homogeneous metals \cite{gotze72}, and to extend its validity using a self-consistent theory for $\Lambda(q,\omega)$ \cite{prelovsek16}. Although approximate, it copes with several nontrivial features of MBL: a) for modest $W>0$ it yields a maximum of 
$\mathrm{Re}~\widetilde{\sigma}(q,\omega)$ at $\omega^* >0$, b) if evaluated at finite effective length $L^*$ it gives a transition to a nonergodic state with $\sigma_0 \sim 0$ and a finite stiffness $S_0(q)>0$, in analogy to self-consistent theories of Anderson localization \cite{gotze79,vollhardt80}. c) due to the coupling to energy fluctuations, i.e., to the the low--$\omega$ energy diffusion mode in the ergodic phase, there is a singular behaviour in 1D when increasing the effective system size $L^* \to \infty$. The latter is an indication that the diffusive behaviour might be unstable in 1D, leading  in the thermodynamic limit $L \to \infty$ to a solution more consistent with a subdiffusion $\sigma(\omega \to 0) \to 0$ \cite{prelovsek16}. 

\section{Dynamical conductivity and d.c. transport}
\label{sec:cond}

The vanishing of the d.c. transport and the absence of diffusion has been the essential novel feature of the Anderson localization \cite{anderson58,mott68,kramer93}, as a direct consequence of the localization of single-particle eigenfunctions. Transport properties have been used as one of criteria for the MBL at finite $T>0$ in the interacting systems \cite{fleishman80,basko06}, being more recently investigated for high $T$ by several numerical studies   \cite{berkelbach10,barisic10,agarwal15,lev15,steinigeweg15,barisic16}. Namely, for large disorders $W>W^*$, one expects that all MB states are localized so that the system should exhibit no d.c. transport at any $T$.  It should be, however, acknowledged that the reverse situation of vanishing d.c. transport does not yet necessarily imply 
the nonergodicity and the full MBL. 

\subsection{Optical conductivity}

The question of dynamical conductivity and d.c. transport is usually posed in the context of uniform charge (or density) current in the 1D chain,
\begin{equation}
j = \imath t_0 \sum_{i} ( c^\dagger_{i+1} c_{i} - c^\dagger_{i} c_{i+i})\,.
\end{equation}
Within the linear response theory the optical/dynamical charge conductivity $\sigma(\omega)$ may be expressed for $T>0$ as
\begin{equation}
\sigma(\omega) = \frac{1- \mathrm{e}^{-\beta \omega}}{L\,\omega} \mathrm{Re}
\int_0^\infty \mathrm{d}t\,\mathrm{e}^{\imath \omega t}
\langle j(t) j \rangle\,.
\label{sigom}
\end{equation}
In the $\beta \to 0$ limit, Eq.~\eqref{sigom} involves a simple averaging over the whole Hilbert space of $N_{\mathrm{st}}$ MB states, 
\begin{equation}
\sigma(\omega)= \frac{\beta \pi}{L N_{\mathrm{st}}}
\sum_{n \neq m} |\langle n |j| m \rangle|^2 \delta(\omega -E_m+ E_n)\,.
\label{cjom}
\end{equation}
Since $\sigma(\omega) \propto \beta$ for large $T$, in this limit it is common to discuss $T \sigma(\omega)$, having the meaning of diffusivity.

\begin{figure}[!htb]
\includegraphics[width=\columnwidth]{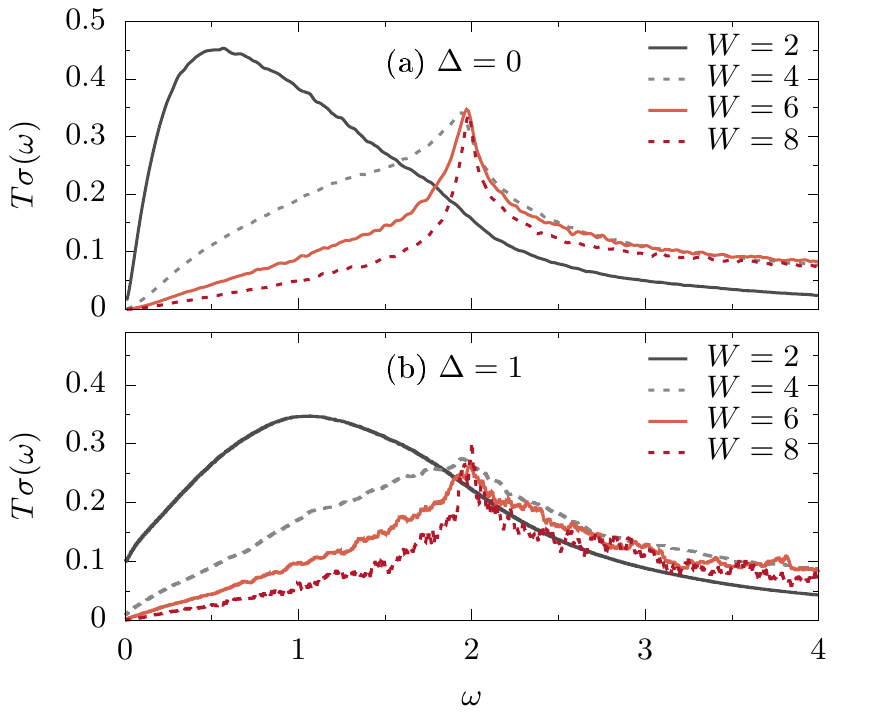}
\caption{\label{fig5} Dynamical conductivity $T \sigma(\omega)$ for various disorders $W=2-8$ and $T\to\infty$: (a) NI $\Delta=0$ (Anderson) model evaluated on a chain with $L=16.000$ sites (exact diagonalization), (b) interacting $\Delta=1$ case as calculated by the MCLM for $L=28$ sites ($\delta\omega \sim 3\times 10^{-3}$ and sampling $N_s \sim 16$).}
\end{figure}
Dynamical conductivity is an intensive macroscopic quantity. For 'normal' ergodic systems, one expects that $\sigma(\omega)$ has a well defined thermodynamic limit $L \to \infty$. However, numerical studies have to deal with finite systems, so the limit $\omega \to 0$ is particularly delicate. That is, in the case of subdiffusive (but ergodic) behaviour \cite{agarwal15,gopal15,znidaric16} the $L \to \infty$ and the $\omega \to 0$ limit do not necessarily commute and $L$ dependence should be analyzed with care. Still, the existence of the nonergodic MBL phase plausibly requires 
vanishing $T \sigma_0 =0$, at least for systems that are large enough $L > l_{loc}$, with $l_{loc}$ playing the role of an effective MB localization length. Again, the order of limits is nontrivial. In order to establish whether $\sigma_0$ is finite or not, one should first consider the limit $L \to \infty$ (check $L$ dependence at $\omega>0$) and then perform the $\omega \to 0$ step.

Let us first discuss some gross features of $T\sigma(\omega)$.
In Fig.~\ref{fig5}, the high-$T$ interacting $\Delta=1$ case is compared with the NI ($\Delta=0$) one \cite{barisic16}. We note that within the NI (Anderson) model the calculation of Eq.~\eqref{cjom} reduces to the search of single-particle eigenstates and corresponding matrix elements, so that the problem can be solved numerically for very large systems, e.g., $L \sim 16.000$ sites in Fig.~\ref{fig5}. 

For $W \geq W^*$, we note from the comparison Fig.~5 that some features of the spectra are very similar, irrespectively of the interaction \cite{barisic10,gopal15,steinigeweg15}. The difference is well visible only in the regime $\omega \ll 1$. While in the NI system $\sigma_0 =0$ for all $W>0$ (Anderson localization), for $\Delta>0$, $T \sigma_0$ becomes strongly suppressed for disorder $W>W^*$, indicating the MBL behaviour.  Furthermore, in both cases, one observes a quite sharp maximum 
at $\omega^* \sim 2t_0$, being a reflection of the bare hopping scale, reappearing for large $W$ in driven systems as well \cite{kozarzewski16}.

The situation changes for a weaker disorder. In particular, as observed from Fig.~\ref{fig5}, for $W=2$ the 
position of the maximum of $\sigma(\omega)$ is clearly interaction dependent, whereas the low--$\omega$ behaviour 
can be generally well represented as
\begin{equation}
\sigma(\omega) \sim \sigma_0 + \xi |\omega|^\alpha . \label{sig}
\end{equation}
Results obtained on finite systems $L = 28$, as in Fig.~\ref{fig5} imply $\sigma_0>0$ for $\Delta>0$ 
and $\alpha \sim 1$ \cite{karahalios09,barisic10,steinigeweg15,barisic16}. Again, we cannot exclude the possibility
that an analysis of much larger systems would be be more consistent with the subdiffusion where $\sigma_0 = 0$ and 
$\alpha <1 $\cite{agarwal15,gopal15,gopal16,luitz16,znidaric16,prelovsek16,luschen16}. In this context,
the relevant quantity is the d.c. dielectric polarizability
\begin{equation}
\chi_d = \frac{1}{\pi} \int_{-\infty}^{\infty}  \frac{\sigma(\omega)} {\omega^2} d\omega. \label{chid}
\end{equation}
Namely, within MBL we are dealing with dielectric where an  electric  field  (along the chain) would induce only a finite
polarization, i.e. $\chi_d <\infty$ which requires according to Eq.~(\ref{sig}) $\sigma_0=0$ and $\alpha>1$. The 
subdiffusive response would  be in this respect anomalous, i.e. in spite of $\sigma_0=0 $ Eq.~(\ref{chid}) implies 
$\chi_d \to \infty$.

Numerical analyses of finite systems are relying on the self-averaging hypothesis, i.e., the averaging over a large number 
$N_s$ of disorder realizations (samples) should represent the behaviour in the thermodynamic limit. Thus, it is important to have under control sample-to-sample (STS) fluctuations too, and not just the mean values of considered quantities. Since Eq.~\eqref{cjom} involves a sum of $R \propto N_{\mathrm{st}}^2$ delta functions, for finite $L$ it is only sensible to discuss $\sigma(\omega)$ using spectral-line broadening, i.e., considering finite frequency bins of width $\eta >0$. Let us denote a broadened result of single disorder realization $k$ by $\sigma^k_\eta(\omega)$ and a sample-averaged spectrum by $\sigma(\omega)=\langle\sigma^k_\eta(\omega) \rangle$. Then, STS fluctuations (that still depend on $\eta$ and $L$) may be characterized by the relative deviation 
$r_{\eta}(\omega)$,
\begin{equation}
r_{\eta}(\omega)= 
\sqrt{\langle [\sigma^k_\eta(\omega) - \sigma(\omega) ]^2\rangle}/\sigma(\omega)\,.
\label{reta}
\end{equation}
Regarding the STS fluctuations, the localized NI systems and the interacting systems exhibit crucial differences. In the localized NI system the single--particle localization with the characteristic length $l^*_{\mathrm{NI}}$ divides the system into $L/l_{\mathrm{NI}}^*$ independent sections. The absence of correlations  (in each sample $k$) between $n$ neighbouring frequency bins lead to a simple scaling $r_{n\eta} = r_{\eta}/\sqrt n$, so that one obtains $r_{\eta}(\omega)\propto 1/\sqrt{\eta L}$, as indeed observed \cite{barisic16}.

Unlike for the localized NI systems, for interacting $\Delta >0$ cases $r_\eta(\omega)$ do not show a significant dependence 
on $\eta$. This indicates that, even for large $W$, the system preserves its MB nature \cite{deutsch91,srednicki94}, according to which the number of different contributions to a given bin in Eq.~\eqref{cjom} is exponentially large in $L$. The latter suppresses the fluctuations up to very narrow bins $\eta$ (beyond the reach of the MCLM resolution $\delta\omega$). With respect to the NI localized case, this leads to a different scaling, $r_{\eta}(\omega) \propto 1/\sqrt{L}$. Yet, for $W> W^*$ actual $r_{\eta}(\omega)$ can be quite considerable even for the largest $L=28$ presented here. Furthermore, for large $W$, the numerical results show a significant $\eta$ dependence in the $\omega\rightarrow0$ limit. Obviously, for $r_\eta(\omega)>1$, the self-averaging hypothesis cannot be directly applied to justify 
the macroscopic relevance of mean values of calculated quantities.
\begin{figure}[!htb]
\includegraphics[width=\columnwidth]{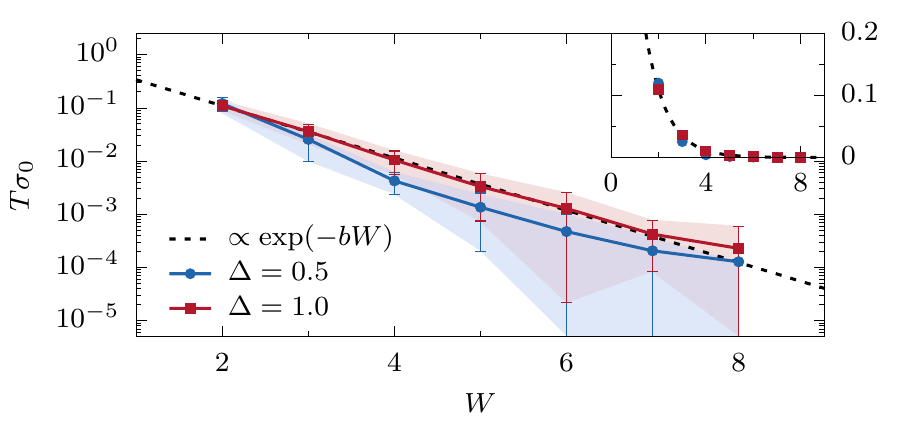}
\caption{\label{fig6} (a) D.c. conductivity $T \sigma_0$ (note log scale) vs. disorder strength $W$, evaluated via MCLM ($T\to\infty$) on $L=28$ sites for $\Delta=0.5,1$, together with the fluctuations (the color area corresponds to sample-to-sample deviations). The inset shows $\sigma_0$ vs. $W$ for $\Delta=1$ in the normal scale fitted to an exponential $\sigma_0 \propto \exp(-bW)$ with $b=1.12$.}
\end{figure}

With these caveats, we analyze and present results for $\sigma_0$ in Eq.~\eqref{sig}. The behaviour of $T \sigma_0$, as
extracted from MCLM results for $L=28$ sites,  and the corresponding STS fluctuations $r_\eta(0)$ may be well observed from Fig.~\ref{fig6}. In particular, for $W \leq W^*$ $\sigma_0$ appears to exhibit a strong exponential-like dependence, $\sigma_0\propto\exp(-bW)$. Furthermore, as seen from Fig.~\ref{fig6}, with increasing $W$ the STS fluctuations of $\sigma_0$ monotonically increase too, reaching $r_{\eta}(0)>1$ values. While on the basis of Fig.~\ref{fig6} it is hard to distinguish between a sharp transition and a crossover to the MBL phase, results are well consistent with other numerical results \cite{steinigeweg15} as well surprisingly close to experimental ones for the mobility of cold atoms \cite{kondov15} in disordered three-dimensional lattice.

\subsection{Thermal conductivity}

Besides the total particle number $N$ (or $S^z_{\mathrm{tot}}$ within the spin representation), for the MBL equally relevant is the other conserved quantity, i.e., the energy. Namely, in order to confirm the MBL in the $\omega\rightarrow0$ limit the 
dynamical thermal conductivity $\kappa(\omega)$ should vanish too,
\begin{equation}
\kappa(\omega)= \beta \frac{1- \mathrm{e}^{-\beta \omega}}{L\,\omega} 
\mathrm{Re} \int_0^\infty \mathrm{d}t\,
\mathrm{e}^{\imath \omega t} \langle j_\epsilon(t) j_\epsilon \rangle.
\label{kappaom}
\end{equation}
Here, $j_\epsilon$ is the energy current operator, defined in a homogeneous as well as in an inhomogeneous system as $j_\epsilon= \sum_{lm} l_r [h_m,h_l]$ \cite{karahalios09}, where $h_l$ are local energy operators following from \eqref{tv}, and $l_r$ corresponding coordinates.

Dynamical thermal conductivity $\kappa(\omega)$ and its d.c. value $\kappa_0 = \kappa(\omega=0)$ have been much
less investigated within the framework of models relevant for MBL \cite{karahalios09,pekker14,varma15}. 
Related question of localization and vanishing of d.c. spin and energy transport has been addressed 
within classical disordered Heiseberg chain \cite{oganesyan09,jencic15},  where there is no signature of 
vanishing d.c. transport. That is, in spite of Anderson-like localization for $T \to 0$ as a common property, 
there seems to be no classical  analogue of the MBL physics.

\begin{figure}[!htb]
\includegraphics[width=\columnwidth]{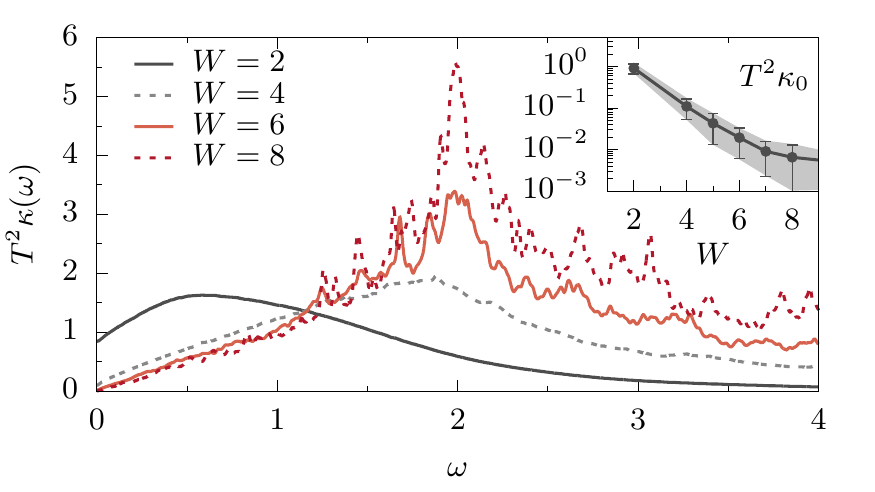}
\caption{\label{fig7} Dynamical thermal conductivity $T^2 \kappa(\omega)$ for various disorders $W=2-8$ for $\Delta=1$ as calculated on system with $L=26$ sites using MCLM at $T\to\infty$. The inset shows d.c. value $T^2\kappa_0$ vs. disorder strength $W$. The color area corresponds to sample-to-sample deviations.}
\end{figure}

Results obtained via MCLM for $L=26$, as presented in Fig.~\ref{fig7}, show that the position of the maximum of $\kappa(\omega)$ for $W>W^*$ is [as for $\sigma(\omega)$], again at $\omega^* = 2t_0$. In contrast to $\sigma(\omega)$, however, the sum rule for $\kappa(\omega)$ is not independent of $W$ \cite{karahalios09}. As shown in the inset of Fig.~\ref{fig7}, the d.c. value $\kappa_0$ is exponentially suppressed by the disorder $W $. It changes by more 
than two order of magnitudes, similarly to $\sigma_0$ in Fig.~\ref{fig6}, apparently with the 
same critical $W^*$ characterizing the MBL transition. 
\section{Disordered Hubbard model}
\label{sec:hubbard}

\subsection{Hubbard chain with potential disorder}

As emphasized already, most theoretical studies concentrated on the disordered 1D model of spinless fermions, \eqref{tv}. However, the cold-atom experiments with fermionic atoms on 1D optical lattice \cite{schreiber15,bordia16,luschen16} 
are realizations of the disordered Hubbard model,
\begin{equation}
H = - t_0 \sum_{is } ( c^\dagger_{i+1,s} c_{is} + \mathrm{h.c.}) + U \sum_i n_{i\uparrow}
n_{i\downarrow} + \sum_i \epsilon_i n_i\,,
\label{hub}
\end{equation}
where $n_j=n_{j \uparrow} + n_{j \downarrow}$ is the local charge (density). Again, we set $t_0=1$. The essential difference with respect to the spinless model, Eq.~\eqref{tv}, is that the Hubbard model \eqref{hub} has two local degrees of freedom: charge (density) $n_i$ and spin (magnetization) $m_i=n_{i \uparrow} - n_{i \downarrow}$. It is important to realize from Eq.~\eqref{hub} that random potential $\epsilon_i$ couples to charge only, 
reflecting the setup in cold-atom experiments. Although in these experiments the potential $\epsilon_i$ is rather quasi-random, 
we proceed with a random uniform distribution $-W < \epsilon_i <W$, 
as usual in the context of Anderson localization.

The disordered Hubbard model has previously attracted less attention\cite{schreiber15,mondaini15,barlev16}, with one conclusion common to the spinless model: the density imbalance appears to be nonergodic for strong disorder, in accordance with experiments \cite{schreiber15,bordia16}. The observation that charge and spin degrees of freedom behave quite differently has been put forward only recently by the present authors \cite{prelovsek216}. This means that within the Hubbard model \eqref{hub} even for strong disorders the system does not follow the full MBL scenario, requiring the existence of a complete set of local conserved quantities~\cite{Nandkishore:2015,huse14,serbyn13}. 

In connection with cold-atom experiments and measured imbalance time evolution $I(t)$, most relevant are charge-density-wave (CDW) correlation functions $C_c(\omega)$ for the particular (staggered) wavevector $q=\pi$. In addition, we study the corresponding spin-density-wave (SDW) correlations $C_s(\omega)$ too,
\begin{eqnarray}
C_c(\omega) &=& \frac{1}{\pi}\mathrm{Re} \int_0^\infty \mathrm{d}t\,
\mathrm{e}^{i \omega^+ t} \langle n_{\pi}(t) n_{\pi} \rangle\,, \nonumber \\
C_s(\omega) &=& \frac{1}{\pi} \mathrm{Re} \int_0^\infty \mathrm{d}t\,
\mathrm{e}^{i \omega^+ t} \langle m_{\pi}(t) m_{\pi} \rangle\,,
\label{eq02}
\end{eqnarray}
where $n_{\pi}= \sum_j (-1)^{j} n_j /\sqrt{L}$, and $m_{\pi}=  \sum_j (-1)^{j} m_j /\sqrt{L}$. In Eq.~\eqref{eq02}, in analogy to the spinless model in Sec.~\ref{sec:model}, the nonergodicity should manifest itself as a singular contribution, $C_c(\omega \sim 0) =C_{c0} \delta(\omega)$, $C_s(\omega \sim 0)=C_{s0} \delta(\omega)$, with $C_{c0}$ and $C_{s0}$ corresponding to the CDW and the SDW stiffnesses, respectively. Obviously, if $C_{c0}$ or/and $C_{s0}$ vanishes, the full MBL is absent.

For calculations of correlations in Hubbard chains we employ again the MCLM, being restricted to $L\leq16$ sites \cite{prelovsek216}. Instead of displaying $C_{c,s}(\omega)$, given by Eq.~\eqref{eq02}, it is more convenient to show 
the 'quasi-time' evolution $\widetilde{C}_{c,s}(\tau)= \int_{-1/\tau}^{1/\tau} \mathrm{d}\omega\,C_{c,s}(\omega)$. The magnetization is fixed to $\bar m = 0$ and the quarter-filled systems $\bar n=1/2$ are investigated, corresponding to experimental set-ups \cite{schreiber15,bordia16,luschen16}.
\begin{figure}[!htb]
\includegraphics[width=\columnwidth]{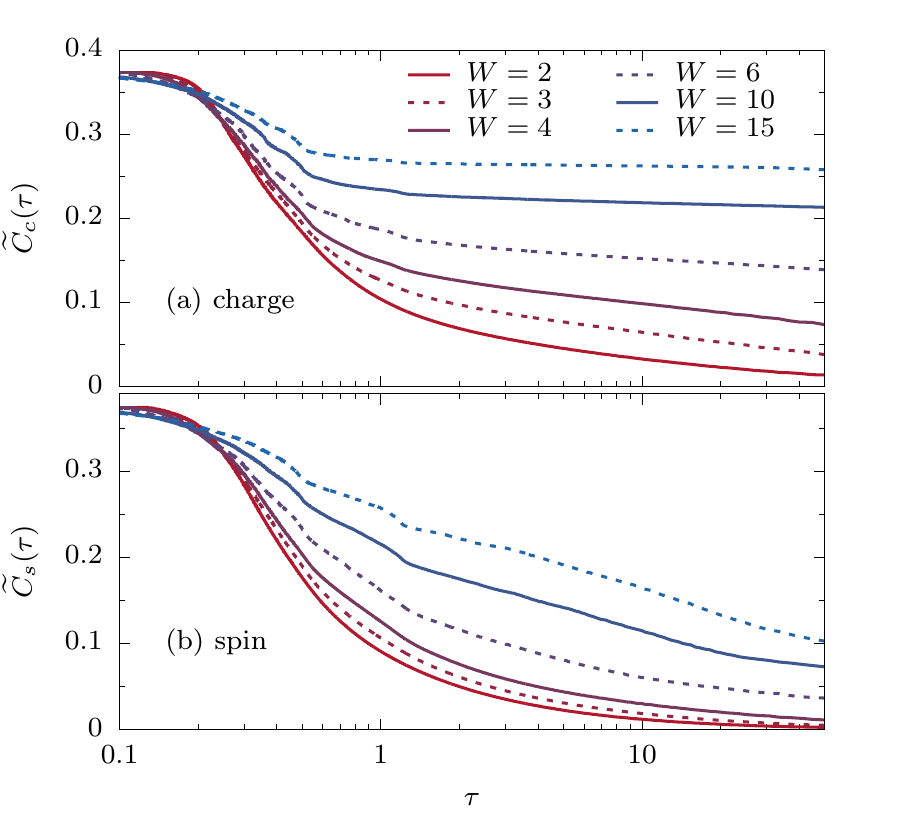}
\caption{\label{fig8} a) CDW correlations $\widetilde{C}_c(\tau)$ vs. 'quasi-time' $\tau$ for quarter-filled ($\bar n= 1/2$) Hubbard model with $U=4$ and various disorders $W = 2-15$. Note the log scale for $\tau$. (b) The same for SDW correlations $\widetilde{C}_s(\tau)$. Results are obtained with MCLM ($T\to\infty$) on the chain of $L=16$ sites.}
\end{figure}

It is plausible that in NI ($U=0$) systems both correlations should be the same $\widetilde{C}_c(\tau)=\widetilde{C}_s(\tau)$, since both spins $n_{i \uparrow}$ and $n_{i \downarrow}$ would localize independently at any $W>0$. But this is not the case for $U>0$. Such a behaviour is clearly observed from Fig.~\ref{fig8} for $U=4$ in a wide window of disorder $W = 2 - 15$. For $\widetilde{C}_c(\tau)$ we observe a behaviour that is qualitatively very similar to the one presented in Fig.~3 for $\widetilde C(\tau)$ within the spinless model \cite{luitz15,mierzejewski16}, and the one reported in experiments \cite{schreiber15,bordia16}. Namely CDW correlations are ergodic $\widetilde{C}_c(\tau \to \infty)=C_{c0} \to 0$ for weak disorders $W \leq W^*$, while for large $W > W^* $ the nonergodicity sets in with $C_{c0} >0$. 

From Fig.~\ref{fig8} one observes an evident contrast with the spin imbalance $\widetilde{C}_s(\tau)$, which appears to decay to zero, $C_{s0} \to 0$, for all considered disorders, i.e., even for the strongest disorder $W =15$. The location of the ergodic-nonergodic transition for CDW correlations in Fig.~\ref{fig8} is harder to fix, being in the range $W^* \sim 4-6$ (for $U=4$). Furthermore, such $W^*$ may be quantitatively compared with the onset of MBL in the 1D optical lattice with the quasi-periodic potential \cite{schreiber15,bordia16,luschen16}. On the other hand, no such transition can be observed for SDW correlations in Fig.~\ref{fig8}, at least not in the investigated range of $W$. It should be mentioned that numerical results for the 
quasi-periodic potential \cite{luschen16} and emerging conclusions do not  differ in any 
essential way from those for the random potential (provided that the disorder $W$ is strong enough \cite{schreiber15}).

An analogous analysis to the dynamical conductivity $\sigma(\omega)$ in Sec.~\ref{sec:cond} can be performed within the Hubbard model for both, charge $j_c$ and spin $j_s$, currents,
\begin{equation}
j_{c,s} = \imath t_0 \sum_{is } [~1,  (\pm 1)^s~] ( c^\dagger_{i+1,s} c_{is}-c^\dagger_{is} c_{i+i,s})\, ,
\end{equation}
by defining corresponding charge and spin conductivities $\sigma_{c,s}(\omega)$, respectively, as in Eq.~\eqref{cjom}. Results \cite{prelovsek216} again confirm the striking difference between charge and spin dynamics:

\noindent a) $\sigma_c(\omega)$ reveals the behaviour very similar to $\sigma(\omega)$ within the spinless model \cite{karahalios09,barisic10,steinigeweg15,barisic16}, as presented in Sec.~\ref{sec:cond}. The maximum of $\sigma_c(\omega)$ at moderate disorder $W\geq 1$ is at $\omega_c^* >0 $, reflecting again the NI limit. At low $\omega \ll1$, $\sigma_c(\omega)$ follows (similarly to $\sigma(\omega)$) a generic nonanalytical behaviour $\sigma_c(\omega) \sim \sigma_c(0)+ \xi |\omega|^\alpha$ with $\alpha \sim 1$. The d.c. value $\sigma_s(0)$ is rapidly vanishing for $W> W^* $.

\noindent b) The behaviour of $\sigma_s(\omega)$ is qualitatively different. First, it exhibits an additional energy scale of spin dynamics, which is well resolved with the maxima of $\sigma_s(\omega)$ at $\omega^*_s \ll 1$, being much lower than $\omega^*_c$ in the charge dynamics. Results \cite{prelovsek216} are consistent with finite $\sigma_s(0) > 0$ up to strongest considered disorder $W = 20$. At the same time, the low-$\omega$ seems to follow $\sigma_s(\omega) \sim \sigma_s(0)+ \tilde \xi |\omega|^{\tilde \alpha}$, but with $\tilde \alpha <1$ even for the largest $W$. This indicates that even if the finite-size scaling would eventually result in $\sigma_s(0) \to 0$, the spin dynamics with $\tilde \alpha<1$ would be ergodic and consistent with a subdiffusive dynamics, as might be the case within the spinless model 
for $W<W^*$ \cite{agarwal15,gopal15,znidaric16}. We note also that $\tilde \alpha<1$, 
together with $\sigma_s(0)=0$, would still 
imply that the static magnetic polarizability (defined in analogy to Eq.~(\ref{chid})) 
$\chi_m = (1/\pi) \int \mathrm{d}\omega\,\sigma_s(\omega)/\omega^2 \to \infty$, i.e., 
the magnetic field along the chain would not induce a finite magnetization, but rather a 
macroscopic shift of the magnetic moment.

Besides the results discussed so far, there are others indicating that the Hubbard chain with a potential disorder, Eq.~\eqref{hub}, does not follow the scenario of the full MBL. In particular, in spite of nonergodicity of the CDW excitation at large $W >W^*$ the entanglement $S_2(t)$, as started from an initial product state and followed via a time-dependent matrix-product method \cite{prelovsek216}, does not increase logarithmically \cite{znidaric08}, but rather follows a power law increase, $S_2(t) \propto t^\mu$ with $0< \mu < 1$. While these observations exclude the full MBL, they open a new class of  nonergodic interacting systems, which could be rationalized as a disordered-induced 
charge-spin separation across all energy scales. 
Nevertheless, it is hard to exclude the possibility that on a very long time scale the spin-charge coupling would lead to thermalization of the whole system \cite{parameswaran16}. 

The essential ingredient to the above striking difference between the charge and spin behaviour is the disorder coupling only to charge, which preserves the SU(2) symmetry. The fundamental role of the latter \cite{potter16,parameswaran16} is not clear, since it can be broken even by, e.g., an homogeneous magnetic field, presumably without any essential effect on results. On the other hand,  an introduction of an additional disorder 
in local random magnetic field (at least if it is strong enough) can evidently induce nonergodicity of SDW correlations \cite{prelovsek216}.

\subsection{Higher dimensional models}

At the moment, there are very few theoretical studies of higher-dimensional models of the MBL \cite{chandran16,hyatt16,deroeck16}, although there exist already several experiments with cold fermions or bosons on 2D and 3D optical lattices, investigating the onset of the MBL \cite{kondov15,bordia16,choi16}. It is evident that numerical investigations of higher-dimensional models involve considerable restrictions, since even in 1D studies finite-size 
and finite-time effects are present. An obvious extension of the 1D spinless model is to (spin) ladder systems \cite{khait16}. Regarding dynamical conductivity $\sigma(\omega)$, the latter reveals similarities with the 1D case.

Due to limitations of numerical studies, analytical approaches are required in order to get some insight into higher-dimensional systems. One recent study in this direction is motivated by cold-atom experiments, in which the 1D chains with identical potential disorder are coupled via an inter-chain hopping \cite{bordia16}. The main observation of this study is that in the regime of 1D nonergodic (MBL) behaviour even a weak inter-chain hopping induces an ergodic (or at least much faster) decay of imbalance $I(t)$. The corresponding theoretical model is the Hubbard model on coupled chains \cite{bordia16,prelovsek216},
\begin{eqnarray}
H &=& \sum_j H_{j} - t^\prime \sum_{ijs }
( c^\dagger_{i,j+1,s} c_{ij,s} + \mathrm{H.c.})\,,
\label{chub}
\end{eqnarray}
where $H_j$ describe 1D Hubbard chains, Eq.~\eqref{hub}. The aim is to show theoretically \cite{prelovsek216} that an inter-chain hopping $t^\prime \neq 0$ in \eqref{chub} qualitatively changes the physics, if disorder is identical in all chains. The starting point for the analysis is that in the NI ($U=0$) case the eigenstates are a product of (longitudinal) localized wave functions and perpendicular plane waves. In particular, one can introduce a basis set that is 
solution of the NI problem,
\begin{equation}
|\phi_{mqs}\rangle = \frac{1}{\sqrt{N}} \sum_{ij} \phi_{mi}
\mathrm{e}^{\imath qj} c^\dagger_{ ijs} |0 \rangle =\varphi^\dagger_{mqs} |0 \rangle\,,
\end{equation}
where NI energies are given by $e_{mq} = e_m + \widetilde{e}_q= e_m - 2t^\prime \cos (q)$, with $e_m$ as energies of 1D 
localized states. The interaction mixes NI eigenstates,
\begin{eqnarray}
H_U = \frac{U}{N}\sum_{\substack{ mm^\prime nn^\prime \\ qkp}}
\chi_{mn}^{m^\prime n^\prime } 
\varphi^\dagger_{n^\prime , k+q \downarrow } \varphi_{nk \downarrow } 
\varphi^\dagger_{m^\prime ,p-q\uparrow} \varphi_{m p\uparrow}\,.
\label{hub2d} 
\end{eqnarray} 

When we consider dynamics of the staggered CDW (imbalance) operator, Eq.~\eqref{eq02}, extended as the sum over all chains $A \propto \sum_{ij} (-1)^i n_{ij}$, we can design a perturbation theory, in analogy to a Fermi-golden rule for the decay rate 
$\Gamma$ of the CDW initial configuration,
\begin{equation}
\Gamma \propto \sum_{\underline{m}} p_{\underline{m} } \sum_{ \underline{n}} 
|\langle \underline{n}|F| \underline{m}\rangle|^2 
\delta (E_{\underline{n}} -E_{\underline{m}}).
\label{gamma}
\end{equation}
Here, $|\underline{m}\rangle$ are MB (Anderson) localized states and $p_{\underline{m}}$ their thermodynamic occupation probability, while $F =[H,A] \sim [H_U,A]$ is the effective force for the CDW decay. 

One can use Eq.~\eqref{gamma} even for a qualitative discussion of the decay within a 1D disordered models, i.e., the case with $t^\prime =0$ (an analogous analysis can be extended to, e.g., the question of the d.c. transport quantities). A nontrivial question arises even in this case, that 
matrix elements $\langle \underline{n}|F| \underline{m}\rangle$ connect (single-particle) states which are close in space (decaying exponentially with distance), while only distant localized states can be close in energy so that $e_m \sim e_n$ in 1D \cite{mott68}, in order to contribute to decay in Eq.~(\ref{gamma}). This emerges as the crucial problem of the role of resonances, having been in the core of single particle localization \cite{anderson58,mott68},
but still mainly open in relation to the existence of the MBL \cite{basko06,imbrie16}. 

However, there is an essential difference, introduced by $t^\prime \ne 0$. The inter-chain coupling leads to a continuous spectrum of overlapping initial and final states, so that matrix elements, Eq.~\eqref{gamma}, can have finite values. At least for a weak perturbation $U<t^\prime $, the evaluation of Eq.~{\eqref{gamma} then reduces to an effective density of decay channels, i.e., with an density of states (at $\beta \to 0$), determined by the conservation of the energy between initial and final configurations, i.e.,
\begin{equation}
e_m - e_n \sim \widetilde{e}_{p-q} + \widetilde{e}_{k+q} - \widetilde{e}_{p} - \widetilde{e}_{k}\,.
\end{equation}
With some further simplifications \cite{prelovsek216} one can then evaluate $\Gamma$ directly using Eq.~\eqref{gamma} and such $\Gamma$ can be then considered at least as an additional channel for the CDW decay due to inter-chain coupling. It is possible to show that such an analysis yields in principle $\Gamma > 0$ for any $t^\prime /t_0 \ne 0$, 
so that for small $U <t^\prime ,t_0 $ one gets $\Gamma \propto U^2$. Also, $\Gamma $ is increasing function of $|t^\prime /t_0|$, although not a simple one \cite{prelovsek216}. It should be stressed that the presented analysis deals with identical chains and cannot be directly extended to independent higher-dimensional disorder. 

\section{Conclusions}
\label{sec:conclu}

With respect to dynamical DW correlations and transport quantities discussed in this review, there is a consistent set of results for the 'standard' 1D disordered spinless models. As shown here, time-dependent DW correlations (relevant to experimentally measured time evolution of imbalance) and dynamical transport properties are closely related quantities. In spite of this unifying picture there are several fundamental issues which remain to be clarified.

\noindent a) In the 1D spinless model the transition to the nonergodic behaviour appears to be sharp. Still, taking into account the limitations of numerical calculations regarding reachable sizes $L$ and times $t_{\mathrm{max}}$ (or small frequencies $\omega_{\mathrm{min}}$) it is hard to distinguish between the well-defined transition and the crossover. Moreover, close inspection of dynamics performed above \cite{mierzejewski16} and also elsewhere \cite{agarwal15,gopal15} indicates that static quantities like the CDW stiffness $S_0(q)$ are not well defined near the transition, since they would have to be measured over exceedingly long time--window. Then, instead of the stiffnesses in this connection one should rather follow the universal dynamics characterized by the dynamical exponents, e.g., the dynamical conductivity exponent $\alpha \sim 1$. 

\noindent b) We have restricted our review to results obtained at high temperatures, $T \to \infty$, where the condition for the existence of the MBL involves the localization of all eigenstates. Since one can expect that in considered interacting models at $T \to 0$ also there is no d.c. transport at any disorder, this opens a question of a possible existence of mobility edges \cite{luitz15} in investigated MB models and related consequences for the $T>0$ density correlations. 

\noindent c) Numerical studies confirm in the presence of finite interactions the existence of an ergodic regime in all models for weak enough 
disorders $W<W^*$.  Still, it is an open question when or whether at all in the hydrodynamic regime $q \to 0,\omega \to 0$ the transport is normal diffusive one, in the sense of the usual diffusion equation with a frequency--independent diffusion constant. Our numerical results for $\sigma(\omega)$ for reachable $L$ \cite{barisic10,steinigeweg15,barisic16} reveal for $W<W^*$ finite $\sigma_0>0$ with a nonanalytical $\omega$-behaviour,
with exponent $\alpha \sim1$. On the other hand, several numerical studies  \cite{agarwal15,gopal15,gopal16,luitz16},
in particular finite-size scaling analysis \cite{znidaric16} as well as the imbalance decay in  cold-atom systems 
\cite{luschen16} are interpreted in terms of a subdiffusive behaviour in the restricted range of the ergodic regime 
$W_{sd} < W<W^*$. Here, $W_{sb}$ would mark the lower onset of subdiffusion. An analytical theory \cite{prelovsek16}, based on the self-consistent treatment of density correlations,  confirms this finding as consequence of low (1D) dimensionality. However, it reveals as well that very large systems,
e.g., $L >100$ are required to establish this delicate phenomenon.

\noindent d) Presented results on the Hubbard model with a potential disorder show that the 1D interacting spinless model is not as generic as one might naively expect. The role of different degrees of freedom and of related symmetries is particularly important for the understanding of the MBL physics. Clearly, this applies even more to higher-dimensional disordered models which have hardly been addressed so far.

\begin{acknowledgements}
P.P. acknowledges the support by the program P1-0044 of the Slovenian Research Agency. M.M. acknowledges support from the 2015/19/B/ST2/02856 project of the Polish National Science Centre. O.S.B. acknowledge the support by Croatian Science Foundation Project No. IP-2016-06-7258 and the QuantiXLie Center of Excellence. J. H. is supported by the U.S. Department of Energy, Office of Basic Energy Sciences, Materials Science and Engineering Division. \end{acknowledgements}

\bibliography{ref_manuadp}

\providecommand{\WileyBibTextsc}{}
\let\textsc\WileyBibTextsc
\providecommand{\othercit}{}
\providecommand{\jr}[1]{#1}
\providecommand{\etal}{~et~al.}


\begin{thebibliography}{[10]}

\bibitem{anderson58}
 \textsc{P.\,W. Anderson} \jr{Phys. Rev.} \textbf{109}, 1492 (1958).


\bibitem{mott68}
 \textsc{N.\,F. Mott} \jr{Phil. Mag.} \textbf{17}, 1259 (1968).


\bibitem{kramer93}
 \textsc{B.~Kramer} and  \textsc{A.~MacKinnon} \jr{Rep. Prog. Phys.}
  \textbf{56}, 1469 (1993).


\bibitem{fleishman80}
 \textsc{L.~Fleishman} and  \textsc{P.\,W. Anderson} \jr{Phys. Rev. B}
  \textbf{21}, 2366 (1980).


\bibitem{basko06}
 \textsc{D.\,M. Basko},  \textsc{I.\,L. Aleiner},  and  \textsc{B.\,L.
  Altshuler} \jr{Ann. Phys.} \textbf{321}, 1126 (2006).


\bibitem{oganesyan07}
 \textsc{V.~Oganesyan} and  \textsc{D.\,A. Huse} \jr{Phys. Rev. B} \textbf{75},
  155111 (2007).


\bibitem{torres15}
 \textsc{E.\,J. Torres-Herrera} and  \textsc{L.\,F. Santos} \jr{Phys. Rev. B}
  \textbf{92}, 014208 (2015).


\bibitem{luitz15}
 \textsc{D.\,J. Luitz},  \textsc{N.~Laflorencie},  and  \textsc{F.~Alet}
  \jr{Phys. Rev. B} \textbf{91}, 081103 (2015).


\bibitem{serbyn16}
 \textsc{M.~Serbyn} and  \textsc{J.\,E. Moore} \jr{Phys. Rev. B} \textbf{93},
  041424(R) (2016).


\bibitem{vasseur16}
 \textsc{R.~Vasseur},  \textsc{A.\,J. Friedman},  \textsc{S.\,A. Parameswaran},
   and  \textsc{A.\,C. Potter} \jr{Phys. Rev. B} \textbf{93}, 134207 (2016).


\bibitem{znidaric08}
 \textsc{M.~\v{Z}nidari\v{c}},  \textsc{T.~Prosen},  and
  \textsc{P.~Prelov\v{s}ek} \jr{Phys. Rev. B} \textbf{77}, 064426 (2008).


\bibitem{bardarson12}
 \textsc{J.\,H. Bardarson},  \textsc{F.~Pollmann},  and  \textsc{J.\,E. Moore}
  \jr{Phys. Rev. Lett.} \textbf{109}, 017202 (2012).


\bibitem{serbyn15}
 \textsc{M.~Serbyn},  \textsc{Z.~Papi{\'{c}}},  and  \textsc{D.\,A. Abanin}
  \jr{Phys. Rev. X} \textbf{5}, 041047 (2015).


\bibitem{berkelbach10}
 \textsc{T.\,C. Berkelbach} and  \textsc{D.\,R. Reichman} \jr{Phys. Rev. B}
  \textbf{81}, 224429 (2010).


\bibitem{barisic10}
 \textsc{O.\,S. Bari\v{s}i\'{c}} and  \textsc{P.~Prelov\v{s}ek} \jr{Phys. Rev.
  B} \textbf{82}, 161106 (2010).


\bibitem{agarwal15}
 \textsc{K.~Agarwal},  \textsc{S.~Gopalakrishnan},  \textsc{M.~Knap},
  \textsc{M.~M\"{u}ller},  and  \textsc{E.~Demler} \jr{Phys. Rev. Lett.}
  \textbf{114}, 160401 (2015).


\bibitem{lev15}
 \textsc{Y.~Bar~Lev},  \textsc{G.~Cohen},  and  \textsc{D.\,R. Reichman}
  \jr{Phys. Rev. Lett.} \textbf{114}, 100601 (2015).


\bibitem{steinigeweg15}
 \textsc{R.~Steinigeweg},  \textsc{J.~Herbrych},  \textsc{F.~Pollmann},  and
  \textsc{W.~Brenig} \jr{Phys. Rev. B} \textbf{94}, 180401(R) (2016).


\bibitem{barisic16}
 \textsc{O.\,S. Bari\v{s}i\'{c}},  \textsc{J.~Kokalj},  \textsc{I.~Balog},  and
   \textsc{P.~Prelov\v{s}ek} \jr{Phys. Rev. B} \textbf{94}, 045126 (2016).


\bibitem{monthus10}
 \textsc{C.~Monthus} and  \textsc{T.~Garel} \jr{Phys. Rev. B} \textbf{81},
  134202 (2010).


\bibitem{pal10}
 \textsc{A.~Pal} and  \textsc{D.\,A. Huse} \jr{Phys. Rev. B} \textbf{82},
  174411 (2010).


\bibitem{luitz16}
 \textsc{D.\,J. Luitz},  \textsc{N.~Laflorencie},  and  \textsc{F.~Alet}
  \jr{Phys. Rev. B} \textbf{93}, 060201 (2015).


\bibitem{mierzejewski16}
 \textsc{M.~Mierzejewski},  \textsc{J.~Herbrych},  and
  \textsc{P.~Prelov{\v{s}}ek} \jr{Phys. Rev. B} \textbf{94}, 224207 (2016).


\bibitem{prelovsek16}
 \textsc{P.~Prelov{\v{s}}ek} and  \textsc{J.~Herbrych} \jr{arXiv:1609.05450}
  (2016).


\bibitem{serbyn13}
 \textsc{M.~Serbyn},  \textsc{Z.~Papi\'{c}},  and  \textsc{D.\,A. Abanin}
  \jr{Phys. Rev. Lett.} \textbf{110}, 260601 (2013).


\bibitem{huse14}
 \textsc{D.\,A. Huse},  \textsc{R.~Nandkishore},  and  \textsc{V.~Oganesyan}
  \jr{Phys. Rev. B} \textbf{90}, 174202 (2014).


\bibitem{rahul15}
 \textsc{N.~Rahul} and  \textsc{D.\,A. Huse} \jr{Annu. Rev. Condens. Matter
  Phys.} \textbf{6}, 15 (2015).


\bibitem{schreiber15}
 \textsc{M.~Schreiber},  \textsc{S.\,S. Hodgman},  \textsc{P.~Bordia},
  \textsc{H.\,P. L\"{u}schen},  \textsc{M.\,H. Fischer},  \textsc{R.~Vosk},
  \textsc{E.~Altman},  \textsc{U.~Schneider},  and  \textsc{I.~Bloch}
  \jr{Science} \textbf{349}, 842 (2015).


\bibitem{kondov15}
 \textsc{S.\,S. Kondov},  \textsc{W.\,R. McGehee},  \textsc{W.~Xu},  and
  \textsc{B.~DeMarco} \jr{Phys. Rev. Lett.} \textbf{114}, 083002 (2015).


\bibitem{bordia16}
 \textsc{P.~Bordia},  \textsc{H.\,P. L\"{u}schen},  \textsc{S.\,S. Hodgman},
  \textsc{M.~Schreiber},  \textsc{I.~Bloch},  and  \textsc{U.~Schneider}
  \jr{Phys. Rev. Lett.} \textbf{116}, 140401 (2016).


\bibitem{choi16}
 \textsc{J.\,Y. Choi},  \textsc{S.~Hild},  \textsc{J.~Zeiher},
  \textsc{P.~Schau{\ss}},  \textsc{A.~Rubio-Abadal},  \textsc{T.~Yefsah},
  \textsc{V.~Khemani},  \textsc{D.\,A. Huse},  \textsc{I.~Bloch},  and
  \textsc{C.~Gross} \jr{Science} \textbf{352}, 1547 (2016).


\bibitem{luschen16}
 \textsc{H.\,P. L{\"{u}}schen},  \textsc{P.~Bordia},  \textsc{S.~Scherg},
  \textsc{F.~Alet},  \textsc{E.~Altman},  \textsc{U.~Schneider},  and
  \textsc{I.~Bloch} \jr{arXiv:1612.07173} (2016).


\bibitem{tsukada99}
 \textsc{I.~Tsukada},  \textsc{Y.~Sasago},  \textsc{K.~Uchinokura},
  \textsc{A.~Zheludev},  \textsc{S.~Maslov},  \textsc{G.~Shirane},
  \textsc{K.~Kakurai},  and  \textsc{E.~Ressouche} \jr{Phys. Rev. B}
  \textbf{60}, 6601 (1999).


\bibitem{yamada01}
 \textsc{T.~Yamada},  \textsc{Z.~Hiroi},  and  \textsc{M.~Takano} \jr{J. Solid
  State Chem.} \textbf{156}, 101 (2001).


\bibitem{shiroka11}
 \textsc{T.~Shiroka},  \textsc{F.~Casola},  \textsc{V.~Glazkov},
  \textsc{A.~Zheludev},  \textsc{K.~Pr\v{s}a},  \textsc{H.\,R. Ott},  and
  \textsc{J.~Mesot} \jr{Phys. Rev. Lett.} \textbf{106}, 137202 (2011).


\bibitem{casola12}
 \textsc{F.~Casola},  \textsc{T.~Shiroka},  \textsc{V.~Glazkov},
  \textsc{A.~Feiguin},  \textsc{G.~Dhalenne},  \textsc{A.~Revcolevschi},
  \textsc{A.~Zheludev},  \textsc{H.\,R. Ott},  and  \textsc{J.~Mesot} \jr{Phys.
  Rev. B} \textbf{186}, 165111 (2012).


\bibitem{thede12}
 \textsc{M.~Thede},  \textsc{F.~Xiao},  \textsc{C.~Baines},
  \textsc{C.~Landee},  \textsc{E.~Morenzoni},  and  \textsc{A.~Zheludev}
  \jr{Phys. Rev. B} \textbf{86}, 180407 (2012).


\bibitem{Mohan2014}
 \textsc{A.~Mohan},  \textsc{N.\,S. Beesetty},  \textsc{N.~Hlubek},
  \textsc{R.~Saint-Martin},  \textsc{A.~Revcolevschi},  \textsc{B.~B\"uchner},
  and  \textsc{C.~Hess} \jr{Phys. Rev. B} \textbf{89}, 104302 (2014).


\bibitem{herbrych13}
 \textsc{J.~Herbrych},  \textsc{J.~Kokalj},  and  \textsc{P.~Prelov\v{s}ek}
  \jr{Phys. Rev. Lett.} \textbf{111}, 147203 (2013).


\bibitem{kokalj15}
 \textsc{J.~Kokalj},  \textsc{J.~Herbrych},  \textsc{A.~Zheludev},  and
  \textsc{P.~Prelov\v{s}ek} \jr{Phys. Rev. B} \textbf{91}, 155147 (2015).


\bibitem{Yu2012}
 \textsc{R.~Yu},  \textsc{L.~Yin},  \textsc{N.\,S. Sullivan},  \textsc{J.\,S.
  Xia},  \textsc{C.~Huan},  \textsc{A.~Paduan-Filho},  \textsc{N.\,F.
  Oliveira~Jr},  \textsc{S.~Haas},  \textsc{A.~Steppke},  \textsc{C.\,F.
  Miclea},  \textsc{F.~Weickert},  \textsc{R.~Movshovich},  \textsc{E.\,D.
  Mun},  \textsc{B.\,L. Scott},  \textsc{V.\,S. Zapf},  and
  \textsc{T.~Roscilde} \jr{Nature} \textbf{489}, 379 (2012).


\bibitem{Wulf2013}
 \textsc{E.~Wulf},  \textsc{D.~H\"{u}vonen},  \textsc{J.\,W. Kim},
  \textsc{A.~Paduan-Filho},  \textsc{E.~Ressouche},  \textsc{S.~Gvasaliya},
  \textsc{V.~Zapf},  and  \textsc{A.~Zheludev} \jr{Phys. Rev. B} \textbf{88},
  174418 (2013).


\bibitem{Povarov2015}
 \textsc{K.\,Y. Povarov},  \textsc{E.~Wulf},  \textsc{D.~H\"{u}vonen},
  \textsc{J.~Ollivier},  \textsc{A.~Paduan-Filho},  and  \textsc{A.~Zheludev}
  \jr{Phys. Rev. B} \textbf{92}, 024429 (2013).


\bibitem{herbrych16}
 \textsc{J.~Herbrych} and  \textsc{J.~Kokalj} \jr{Phys. Rev. B} \textbf{95},
  125129 (2017).


\bibitem{bordai16}
 \textsc{P.~Bordia},  \textsc{H.\,P. L\"{u}schen},  \textsc{S.\,S. Hodgman},
  \textsc{M.~Schreiber},  \textsc{I.~Bloch},  and  \textsc{U.~Schneider}
  \jr{Phys. Rev. Lett.} \textbf{116}, 140401 (2016).


\bibitem{torres16}
 \textsc{E.\,J. Torres-Herrera} and  \textsc{L.\,F. Santos} \jr{Ann. Phys.
  (Berlin)} p.\,1600284 (2017).


\bibitem{deutsch91}
 \textsc{J.\,M. Deutsch} \jr{Phys. Rev. A} \textbf{43}, 2046 (1991).


\bibitem{srednicki94}
 \textsc{M.~Srednicki} \jr{Phys. Rev. E} \textbf{50}, 888 (1994).


\bibitem{steinigeweg13}
 \textsc{R.~Steinigeweg},  \textsc{J.~Herbrych},  and
  \textsc{P.~Prelov\v{s}ek} \jr{Phys. Rev. E} \textbf{87}, 012118 (2013).


\bibitem{steinigeweg14}
 \textsc{R.~Steinigeweg},  \textsc{A.~Khodja},  \textsc{H.~Niemeyer},
  \textsc{C.~Gogolin},  and  \textsc{J.~Gemmer} \jr{Phys. Rev. Lett.}
  \textbf{112}, 130403 (2014).


\bibitem{long03}
 \textsc{M.\,W. Long},  \textsc{P.~Prelov\v{s}ek},  \textsc{S.~El~Shawish},
  \textsc{J.~Karadamoglou},  and  \textsc{X.~Zotos} \jr{Phys. Rev. B}
  \textbf{68}, 235106 (2003).


\othercit
\bibitem{prelovsek13}
 \textsc{P.~Prelov\v{s}ek} and  \textsc{J.~Bon\v{c}a},
Ground state and finite temperature lanczos methods,
 in: Strongly Correlated Systems - Numerical Methods, edited by A.~Avella and
  F.~Mancini,  (Springer, Berlin, 2013).


\bibitem{herbrych12}
 \textsc{J.~Herbrych},  \textsc{R.~Steinigeweg},  and
  \textsc{P.~Prelov\v{s}ek} \jr{Phys. Rev. B} \textbf{86}, 115106 (2012).


\bibitem{kozarzewski16}
 \textsc{M.~Kozarzewski},  \textsc{P.~Prelov\v{s}ek},  and
  \textsc{M.~Mierzejewski} \jr{Phys. Rev. B} \textbf{93}, 235151 (2016).


\bibitem{gopal15}
 \textsc{S.~Gopalakrishnan},  \textsc{M.~M\"{u}ller},  \textsc{V.~Khemani},
  \textsc{M.~Knap},  \textsc{E.~Demler},  and  \textsc{D.\,A. Huse} \jr{Phys.
  Rev. B} \textbf{92}, 104202 (2015).


\bibitem{gopal16}
 \textsc{S.~Gopalakrishnan},  \textsc{K.~Agarwal},  \textsc{E.\,A. Demler},
  \textsc{D.\,A. Huse},  and  \textsc{M.~Knap} \jr{Phys. Rev. B} \textbf{93},
  134206 (2016).


\bibitem{znidaric16}
 \textsc{M.~\v{Z}nidari\v{c}},  \textsc{A.~Scardicchio},  and  \textsc{V.\,K.
  Varma} \jr{Phys. Rev. Lett.} \textbf{117}, 040601 (2016).


\othercit
\bibitem{forster95}
 \textsc{D.~Forster},
Hydrodynamic Fluctuations, Broken Symmetry, And Correlation Functions (Westview
  Press, New York, 1995).


\bibitem{potter15}
 \textsc{A.\,C. Potter},  \textsc{R.~Vasseur},  and  \textsc{S.\,A.
  Parameswaran} \jr{Phys. Rev. X} \textbf{5}, 031033 (2015).


\bibitem{gotze72}
 \textsc{W.~G\"{o}tze} and  \textsc{P.~W\"{o}lfle} \jr{Phys. Rev. B}
  \textbf{6}, 1226 (1972).


\bibitem{gotze79}
 \textsc{W.~G\"{o}tze} \jr{J. Phys. C: Solid State Phys.} \textbf{12}, 1279
  (1979).


\bibitem{vollhardt80}
 \textsc{D.~Vollhardt} and  \textsc{P.~W\"{o}lfle} \jr{Phys. Rev. Lett.}
  \textbf{45}, 842 (1980).


\bibitem{karahalios09}
 \textsc{A.~Karahalios},  \textsc{A.~Metavitsiadis},  \textsc{X.~Zotos},
  \textsc{A.~Gorczyca},  and  \textsc{P.~Prelov\v{s}ek} \jr{Phys. Rev. B}
  \textbf{79}, 024425 (2009).


\bibitem{pekker14}
 \textsc{D.~Pekker},  \textsc{G.~Refael},  \textsc{E.~Altman},
  \textsc{E.~Demler},  and  \textsc{V.~Oganesyan} \jr{Phys. Rev. X} \textbf{4},
  011052 (2014).


\bibitem{varma15}
 \textsc{V.\,K. Varma},  \textsc{A.~Lerose},  \textsc{F.~Pietracaprina},
  \textsc{J.~Goold},  and  \textsc{A.~Scardicchio} \jr{arXiv:1511.09144}
  (2015).


\bibitem{oganesyan09}
 \textsc{V.~Oganesyan},  \textsc{A.~Pal},  and  \textsc{D.~Huse} \jr{Phys. Rev.
  B} \textbf{80}, 115104 (2009).


\bibitem{jencic15}
 \textsc{B.~Jen{\v{c}}i{\v{c}}} and  \textsc{P.~Prelov{\v{s}}ek} \jr{Phys. Rev.
  B} \textbf{92}, 134305 (2015).


\bibitem{mondaini15}
 \textsc{R.~Mondaini} and  \textsc{M.~Rigol} \jr{Phys. Rev. A} \textbf{92},
  041601(R) (2015).


\bibitem{barlev16}
 \textsc{Y.~Bar~Lev} and  \textsc{D.\,R. Reichman} \jr{EPL (Europhysics
  Letters)} \textbf{113}, 46001 (2016).


\bibitem{prelovsek216}
 \textsc{P.~Prelov{\v{s}}ek},  \textsc{O.\,S. Bari{\v{s}}i{\'{c}}},  and
  \textsc{M.~{\v{Z}}nidari{\v{c}}} \jr{Phys. Rev. B} \textbf{94}, 241104
  (2016).


\bibitem{Nandkishore:2015}
 \textsc{R.~Nandkishore} and  \textsc{D.\,A. Huse} \jr{Annu. Rev. Condens.
  Matter Phys.} \textbf{6}, 15 (2015).


\bibitem{parameswaran16}
 \textsc{S.\,A. Parameswaran} and  \textsc{S.~Gopalakrishnan} \jr{Phys. Rev. B}
  \textbf{95}, 024201 (2016).


\bibitem{potter16}
 \textsc{A.\,C. Potter} and  \textsc{R.~Vasseur} \jr{Phys. Rev. B} \textbf{94},
  224206 (2016).


\bibitem{chandran16}
 \textsc{A.~Chandran},  \textsc{A.~Pal},  \textsc{C.\,R. Laumann},  and
  \textsc{A.~Scardicchio} \jr{Phys. Rev. B} \textbf{94}, 144203 (2016).


\bibitem{hyatt16}
 \textsc{K.~Hyatt},  \textsc{J.\,R. Garrison},  \textsc{A.\,C. Potter},  and
  \textsc{B.~Bauer} \jr{Phys. Rev. B} \textbf{95}, 035132 (2016).


\bibitem{deroeck16}
 \textsc{W.~De~Roeck} and  \textsc{F.~Huveneers} \jr{arXiv:1608.01815} (2016).


\bibitem{khait16}
 \textsc{I.~Khait},  \textsc{S.~Gazit},  \textsc{N.\,Y. Yao},  and
  \textsc{A.~Auerbach} \jr{Phys. Rev. B} \textbf{93}, 224205 (2016).


\bibitem{imbrie16}
 \textsc{J.\,Z. Imbrie} \jr{Phys. Rev. Lett.} \textbf{117}, 027201 (2016).


\end{thebibliography}
\end{document}